\documentclass[preprint,12pt]{elsarticle}

\usepackage{amssymb}
\usepackage{caption}
\usepackage{subcaption}

\journal{arXiv}

\begin{document}

\begin{frontmatter}



\title{Osteosarcoma Tumor Detection using Transfer Learning Models}


\author[inst1]{Raisa Fairooz Meem}

\affiliation[inst1]{organization={Department of Computer Science, American International University Bangladesh},
            addressline={Dhaka}, 
            city={Dhaka},
            country={Bangladesh}}

\author[inst2]{Khandaker Tabin Hasan}

\affiliation[inst2]{organization={Department of Computer Science, American International University Bangladesh},
            addressline={Dhaka}, 
            city={Dhaka},
            country={Bangladesh}, 
            email= { tabin@aiub.edu}}

\begin{abstract}
The field of clinical image analysis has been applying transfer learning models increasingly due to their less computational complexity, better accuracy etc. These are pre-trained models that don't require to be trained from scratch which eliminates the necessity of large datasets. Transfer learning models are mostly used for the analysis of brain, breast, or lung images but other sectors such as bone marrow cell detection or bone cancer detection can also benefit from using transfer learning models, especially considering the lack of available large datasets for these tasks. This paper studies the performance of several transfer learning models for osteosarcoma tumour detection. Osteosarcoma is a type of bone cancer mostly found in the cells of the long bones of the body. The dataset consists of H\&E stained images divided into 4 categories- Viable Tumor, Non-viable Tumor, Non-Tumor and Viable Non-viable. Both datasets were randomly divided into train and test sets following an 80-20 ratio. 80\% was used for training and 20\% for test. 4 models are considered for comparison- EfficientNetB7, InceptionResNetV2, NasNetLarge and ResNet50. All these models are pre-trained on ImageNet. According to the result, InceptionResNetV2 achieved the highest accuracy (93.29\%), followed by NasNetLarge (90.91\%), ResNet50 (89.83\%) and EfficientNetB7 (62.77\%). It also had the highest precision (0.8658) and recall (0.8658) values among the 4 models.
\end{abstract}

\begin{graphicalabstract}
\includegraphics{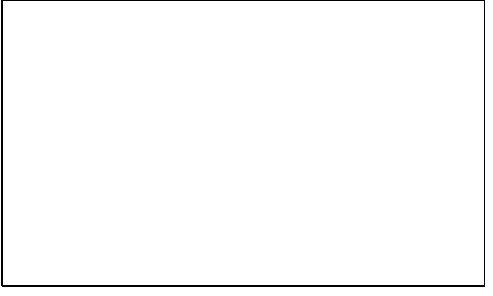}
\end{graphicalabstract}

\begin{highlights}
\item Explore the applications of transfer learning for Ostersarcoma Tumor Detection
\item Conduct a comparative study on the performance of different transfer learning models for Osteosarcoma Tumor Detection
\end{highlights}

\begin{keyword}
Osteosarcoma \sep Tumor Detection \sep Transfer Learning \sep ResNet50 \sep EfficientNet \sep InceptionREsNet \sep NasNetLarge
\PACS 0000 \sep 1111
\MSC 0000 \sep 1111
\end{keyword}

\end{frontmatter}


\section{Introduction}

Medical image analysis is the process of identifying and categorizing particular elements or characteristics in an image \cite{Du2020}. It involves computer-based image processing techniques to classify, extract features, reconstruct and present different organs or tissues which are suspected to be affected by various diseases. The images can be 2D or 3D, and based on the similarities and dissimilarities, experts can detect the problem, considerably enhancing the precision and dependability of medical diagnosis \cite{Liu2021}. It plays a major role in clinical diagnosis and detecting diseases correctly to help medical professionals provide proper treatment to the patient. There are several useful evaluation criteria in this process to ensure accurate detection such as recall, sensitivity, precision, specificity, F-measure etc. The most frequently applied tasks are visualization, segmentation, enhancement etc. \cite{Du2020} The advantage of the image segmentation technique is the facility of remote image analysis. It reduces the computational complexity, time, and cost by eliminating the necessity of biological samples for segmentation \cite{KK2020}.

Osteosarcoma is the most found malignant bone cancer and the earliest hominin cancer to be found around the world \cite{Meltzer2021}. It is the 3rd most commonly found malignant cancer in Bangladesh \cite{Kazi2018}. Adolescents and people aged from 10 to 30 have the highest chance of developing an osteosarcoma tumour, most frequently in the longitudinal bone development areas in the body such as the Distal femur and Proximal Tibia (Knee) or Proximal Humerus (Shoulder). People above 60 have the 2nd highest chance which is frequently linked to Paget's disease of the bone and likely indicates a different biologic process. The average survival rate is 5 years for 60\% of the patients but only 20\% for patients with metastatic diseases found in the lung \cite{Meltzer2021}.

Osteosarcoma diagnosis relies on the examination of radiographic images of affected areas. The degree of tissues and bones being affected by osteosarcoma can be determined using different cross-sectional imaging approaches. CT scan and MRI images are the most used techniques for this task though MRI is more frequently used due to the clearer presentation of conditions such as soft tissue extension, localized intramedullary metastases, and intramedullary beating metastases etc. \cite{Wu2022} After detecting the tumour, doctors can improve the quality of treatment by assessing tumour response to chemotherapy. Tumour necrosis has been playing a significant role in high-grade osteosarcoma treatment for a long time \cite{Babu2019}.

The possibility of survival increases if osteosarcoma is detected at early stages and the patient is monitored diligently but in developing countries like Bangladesh where there is a shortage of qualified radiologists, the medical sector often fails to provide this facility to the patients. In addition, the detection of osteosarcoma is a difficult and time-consuming process that includes cancer and necrosis cell grading during the treatment phase and often requires multiple radiologists involved \cite{Nabid2020}. Moreover, the accuracy of detection also depends on the experience and expertise of the people involved, so, a computerised approach would be preferable to manual detection. There are several factors that might affect the output. Firstly, tumours vary in shape, size, and structure. Secondly, the highly heterogeneous nature of this type of cancer adds more complexity to the process as uneven density distribution between the tumour cell and normal cell makes it hard to differentiate \cite{Rui2018}.

Due to the complexity and difficulty of osteosarcoma detection, medical professionals now prefer computerized diagnosis approaches that not only reduce the complexity but also provide the result (in this case, a tumour being malignant or benign) quickly with impressive accuracy. The approaches typically utilize a feature extractor method to analyze the pre-processed images with enhanced quality and reduced noise along with a classifier. Feature extraction, a process applied to understand data by lowering the dimensionality of the given data, is immensely helpful for data analysis but due to the complications of the process, they are hard to use when complex images are involved. To reduce the complexity, Deep Learning (DL) models are now widely used as feature extractors as they employ faster compact processors along with Convolutional Neural Networks (CNN). They also reduce the requirement of image pre-processing to an extent but the necessity of large datasets for training to avoid overfitting is a major disadvantage of the approach \cite{Nasir2022}.

Traditionally, mathematical approaches along with various types of filters such as edge detection were applied to detect the disease that almost went extinct after the introduction of machine learning which extracted manually created characteristics. However, the complex process of feature extraction and design was considered a drawback which led to the widespread use of Deep learning in this sector \cite{Hesam2019}.

Transfer Learning (TL) is a branch of Deep Learning where the models are pre-trained to perform a task in a way that they can later be applied to perform a different yet related task. This method is lately gaining attention of the researchers interested in computerized medical image analysis as this variety of CNN applies previously initialized weights that can provide significant accuracy if trained with a large dataset. It has the capability to utilize weights used to categorize a dataset to categorize another dataset instead of building the same model over and over \cite{Kandel2020}. ImageNet database is a popular method of training transfer learning models as it provides a publicly available collection of over 14 million manually annotated images among which around one million images contain bounding boxes. It alludes to a circumstance in which what has been discovered in one environment is used to enhance optimization in another one. It is the process of applying previously acquired information from one domain to another for feature extraction and categorization. The process is performed using a deep CNN model that has previously been trained on a sizable dataset. A new dataset with fewer training images than the previously trained datasets is used to further train (fine-tune) the CNN model. The layers at the beginning of CNN models often learn features like edges, curves, corners, etc., whereas the layers at the end focus on more abstract features. Often, the fully connected layer, SoftMax layer, and classification output layer are discharged, and the remaining layers are moved to perform a new classification \cite{Deniz2018}. This eliminates the requirement of large datasets to train the model with randomly initialized weights from scratch which made Transfer Learning a typical choice when the available dataset is small \cite{Huss2018}. The concept of Transfer learning is explained in Figure 1:

\begin{figure}[h] 
\centering
\includegraphics[width=8cm]{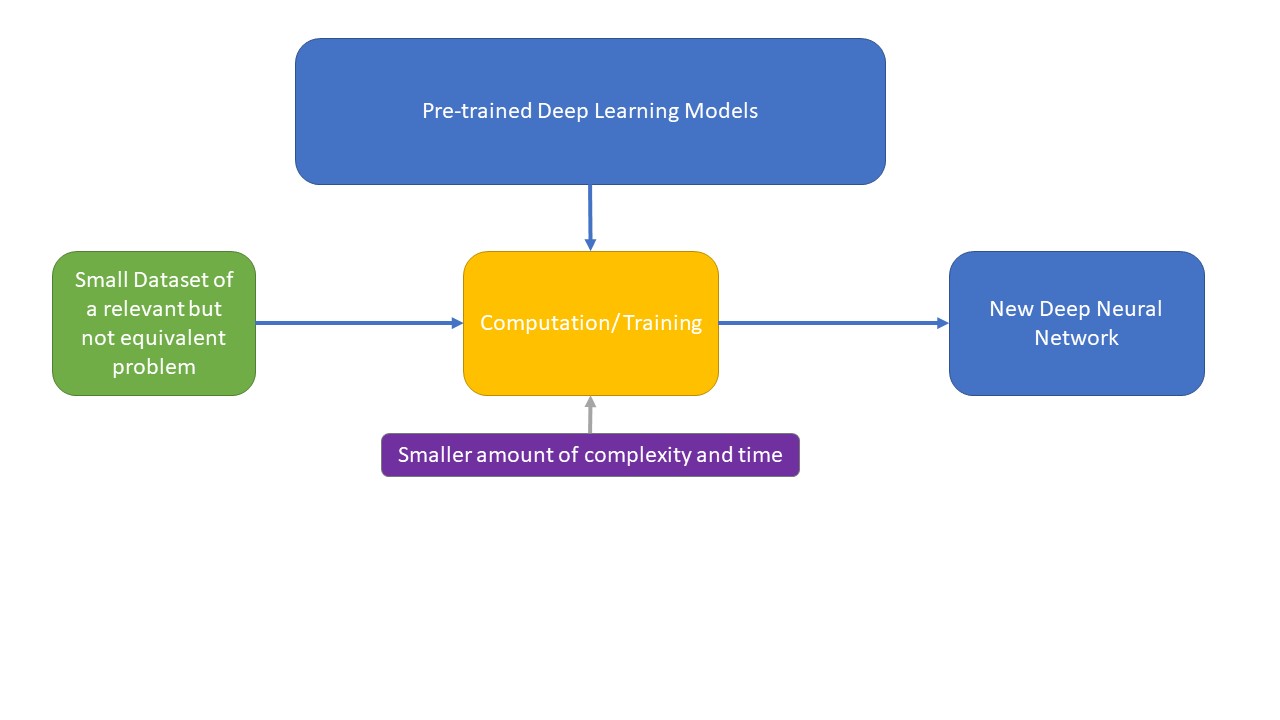}
\caption{Transfer Learning Architecture}
\end{figure}

In this paper, 4 pre-trained TL models (EfficientNet, InceptionResNet, NasNet and ResNet) are applied to a small dataset of Osteosarcoma to detect Osteosarcoma tumours. The models are selected based on their performance on relevant tasks described in the later section. The paper is divided into 5 sections- Section II describes the relevant progress in this domain to date, Section III explains the methods and materials used in this study, Section IV analyses the results and finally a conclusion is provided with future research opportunities in Section V.

\section{Literature Review}
Deep Learning (DL) and Convolutional Neural Networks (CNN) are widely used for medical image segmentation for years. Naturally, the use of Transfer Learning (TL) is also encouraged in this domain, even more so as it offers a solution to the requirement of large datasets for training.

Morid et al. \cite{Morid2021} conducted a scoping review to study different transfer learning models trained on the ImageNet dataset that has been used for medical image analysis where the studied articles were published after 2018 to provide the most updated status of this field. They researched the most used transfer learning models and stated that different models are most customarily used for processing different types of images. For example, Inception models are most frequently used for X-ray, endoscopic, or ultrasound images whereas VGGNet shows comparatively better performance for OCT or skin lesion images. When considering the most frequently used models overall, regardless of image or organ type, the top 4 include Inception, VGGNet, AlexNet, and ResNet. The InceptionResNet model was mentioned in only 3 studies due to being a comparatively newer approach. The preference for Inception and ResNet models was also proved in the literature review conducted by Kim, Hee E., et al. \cite{Kim2022}. 

Shah, Hasnain Ali, et al. \cite{Shah2022} proposed an approach using the transfer learning model EfficientNet for Brain Tumor Detection where the dataset contains MRI images. The proposed method used the EfficientNetB0 model pre-trained with the ImageNet dataset which was developed using a composite scaling technique where the network depth is proportionate to the number of layers in the network. They also used 5 other transfer learning models- VGG16, InceptionV3, Xception, ResNet50 and InceptionResNetV2 for comparing the outputs. A set of specified coefficients are used to adjust each dimension that helped the EfficientNet model outperform other Deep CNN models. The suggested approach uses an optimizer algorithm to modify the biases and learning rate of the neural network which increases accuracy while reducing overall loss. The loss function slowly learns to lower the prediction error with the aid of an optimization function. It provided 98.87\% accuracy for the detection. 

Marques et al. \cite{Marq2020} also used EfficientNet, more specifically EfficientNetB4 for the two tasks of classification of COVID-19 medical images- a. binary classification utilizing photos of COVID-19 patients and healthy patients and b. a discussion of the multi-class results utilizing photos from COVID-19, pneumonia, and healthy patients. The proposed method introduced a "global\_average\_pooling2d" layer to lessen overfitting by limiting the number of parameters. A set of three inner dense layers with RELU activation functions and dropout layers has also been implemented along with a random dropout rate of 30\% to avoid overfitting. To build the suggested automatic detection system, a SoftMax activation function has been introduced to one output dense layer, which has three output units for multiclass classification and two output units for binary classification. The average accuracy of the model for binary and multiclass is 99.62\%, and for binary, it is 96.70\%. In terms of binary and multi-class classification, the EfficientNetB4 model has an average recall value of 99.63\% and 96.69\%, respectively. On the other hand, binary classification has an average precision value of 99.64\%, while multi-class reports a value of 97.54\%. The average F1 score for binary classification is 99.62\% and 97.11\% for multi-class classification. The authors mentioned a few limitations such as the comparison being restricted by the variations in the samples utilized and the machine learning approach parameters. Also, due to the lack of available details in the literature, such as the software used, comparing the methodologies used by the studies that are available using the samples from their research, or vice versa, was not possible.

Falconi et al. \cite{Fal2019} experimented to study the accuracy provided by MobileNet and NasNet for the classification of Breast Mammogram Abnormalities. They claimed to be the 1st ones to use the models MobileNet and NasNet for mammogram images. They created two sub-datasets from the original one- the Otsu Dataset where Otsu's algorithm was applied for image segmentation and the ROI dataset containing tissue information. To compare the accuracy, InceptionV3 and ResNet50 were also used. For the Otsu dataset, NasNet (68.0\%) and InceptionV3 (67.5\%) provided the best accuracy whereas, for the ROI dataset, ResNet50 (78.4\%) and MobileNet (74.3\%) gave the best output. The overall performance of all the models for the ROI dataset was comparatively better than the Otsu dataset as InceptionV3, the model providing the least accuracy also provided an output of 68.9\% which is higher than the maximum accuracy derived for the Otsu dataset. However, they didn't use data augmentation in their experiment. 

Faruk, Omar, et al. \cite{Faruk2021} conducted a comparative study on 4 transfer learning approaches- Xception, InceptionV3, InceptionResNetV2 and MobileNetV2 for tuberculosis detection from X-ray images. Each model consists of a flattened layer, two dense layers, and a ReLU activation function along with three MaxPooling2D levels and four Conv2D layers. The final and thickest layer, SoftMax, acts as an activation layer with a few adjustments in the final layers. Results are tailored using layers such as average pooling, flattening, dense, and drop out. Among the 4 models, InceptionResNetV2 achieved the highest accuracy (99.36\%).

InceptionResNetV2 was also the preferred method for X-ray image classification of the chest to detect pneumonia in the work of Demir, Ahmet, and Feyza Yilmaz \cite{Dem2020}. They made several changes in the model such as the ReLU activation being replaced by a new and improved function called LeakyReLu activation, Averagepooling layers being used instead of Maxpooling etc. These changes were done separately to observe the change in output and then combined in one network for the main experiment. So, there were 4 models technically- InceptionResNetV2, InceptionResNetV2 with LeakyRelu, InceptionResNetV2 with Averagepooling, and the proposed model InceptionResNetV2 with both LeakyReLu and Averagepooling. The proposed model provided 93.59\% specificity with 93.16\% sensitivity. However, it was found that InceptionResNetV2 with LeakyReLu only has better accuracy and specificity than InceptionResNetV2 with both LeakyRelu and Averagepooling. Also, InceptionResNetV2 with Averagepooling only has higher sensitivity than InceptionResNetV2 with both LeakyRelu and Averagepooling. The traditional InceptionResNetV2 had the poorest performance of all 4.

Anisuzzaman, D. M., et al. \cite{Anis2021} performed a study of deep learning models, namely VGG19 and InceptionV3 to detect Osteosarcoma using histological images, claiming this was the 1st use of the said models for osteosarcoma detection. For the VGG19 model, the output layer and completely connected layer of the VGG19 model were removed, and two fully connected layers were added after the final "MaxPool" layer. Along with "Relu" activation in the dense layers and the "SoftMax" activation function in the output layer, dropout layers are utilized to prevent over-fitting the training data. Binary classification and multiclass classification were the two methods used to conduct the studies. VGG19 provided the highest accuracy for both tasks- 95.6\% for binary classifications and 93.9\% for multiclass classifications. Furthermore, the Necrotic Tumor versus Non-Tumor has the highest F1 score in the binary class (0.97).

Mahore et al. \cite{Mahore2021} proposed a Random Forest machine learning algorithm for osteosarcoma detection where they used the most frequently used deep learning models such as VGGNet, CNN, AlexNet, and LeNet for a comparative study. Expert-guided features and Cell-profiler features were recovered from the OST Images. Total clusters, red and blue count, red and blue percentage, Circularity, and Area are expert-guided features. In contrast, Cell-Profiler offers extracts from Cell-profiler Software that contain Gabor, Variance, and Entropy. Osteosarcoma was divided into three categories, namely, viable, non-viable, and non-tumour, using both types of features. The proposed method outperformed the other models by achieving 92.4

Nasir, Muhammad Umar, et al. \cite{Nasir2022} proposed a transfer learning model for Automatic Osteosarcoma Cancer Detection where they added blockchain technology along with fog and edge computing with the model to ensure higher accuracy. The paradigm includes 5 layers- data, pre-processing, edge computing, fog computing, and finally a testing layer. The process begins with data collection using IoMT technology and depositing them in a private data cloud secured by blockchain technology to ensure confidentiality. The edge computing layer utilizes the AlexNet method along with SGDM, ADAM and RMSProp for training data while the fog computing layer optimizes the model further. A huge dataset of WSIs was used in experiments, and the accuracy rate reached up to 99.3

Sarwinda, Devvi, et al \cite{Sarwinda2021} compared two variants of ResNet- ResNet-18 and ResNet-50 for benign and malignant Colorectal Cancer Detection. They took different portions of a dataset to analyse the performance of each model. The experiment's initial pre-processing step makes use of grayscale photos and contrast amplification with CLAHE and applies to learning with the ResNet model. Residual blocks divide the convolution layer into five layers in each ResNet design. Only the initial stage of feature learning, which comes after the first convolution, and the last stage of feature learning, which comes after the prior convolution, is applied before the classification layer is added. The 20\% and 25\% test sets with a classification accuracy of over 80\%, a sensitivity of above 87\%, and a specificity of above 83\% are where we see the best performance value throughout the three test assortments. ResNet50 provided better performance than ResNet80 in all cases in terms of accuracy, sensitivity, and specificity. 

ResNet combined with Deep Convolutional Generative Adversarial Network (DC-GAN) was applied to classify blood cells in the study of Ma, Li, et al. \cite{Ma2020}. To improve the performance of the proposed model, a state-of-the-art loss function was introduced which follows a strategy of hard sample mining for easy filtering of samples. It also reduces the value of intraclass distance to make the interclass distance value higher with an additional margin for better discriminative power of the model. DC-GAN is applied for sample expansion that is used for training the model. For comparison with the proposed model, several other transfer learning models, namely Inception, ResNet50, InceptionV3-LSTM, ResNet50-LSTM, and Xceotion-LSTM were used. The proposed model outperformed all these models by achieving 91.68\% accuracy. The authors also claimed that the proposed model can work perfectly with small datasets or if there is some anomaly in the dataset such as missing labels, imbalance, poor utilization of features etc.

In conclusion, the use of TL models is most frequent in the sector of brain and lung disease detection, especially various types of cancer. However, their performance encouraged researchers to apply them for the detection of diseases in other body organs too. Several models have achieved impressive accuracy for osteosarcoma detection which gives us the motivation to research further into this domain.

\section{Methodology}
\subsection{Dataset}
The image collection is the work of Leavy et al. \cite{P2019} that consists of histology photos of osteosarcomas stained with haematoxylin and eosin (H\&E). Haematoxylin and Eosin staining is often used in cancer research where, in the image, the tissues are dyed pink and nuclei blue. To this date, the procedure requires a manual microscopic examination of stained slides done by pathologists. Detection of cancer, tumour, its size etc. is a laborious act \cite{Mishra2018}.
In the case of osteosarcoma, both normal and tumour cells are stained blue, but the difference lies in the shapes of those cells. Normal cells have rounder and regular shapes whereas irregular shapes indicate tumour cells which also vary in type. This intra-tumour histological variability is so high for osteosarcoma that many other methods designed for other tumour types don’t work for it \cite{Mishra2018}.
A group of clinical scientists at the University of Texas Southwestern Medical Center in Dallas gathered the data. This dataset was produced from archived samples of 50 patients who received care at Children's Medical Center, Dallas, between 1995 and 2015. Pathologists chose four patients (out of 50) based on the variety of tumour specimens found following surgical excision. Two doctors worked together to annotate the text. Two pathologists each received a set of photos for the annotation activity. Since only one pathologist annotated any given image, each image only had one annotation. The collection comprises 1144 photos with a resolution of 10X and a size of 1024 X 1024. The images are labelled into 3 categories- Viable Tumor, Non-viable Tumor and Non-Tumor. An example of images from these 3 classes is shown in Figure 2.

\begin{figure}[h] 
\centering
\begin{subfigure} [b]{0.3\textwidth}
\includegraphics[width=35mm]{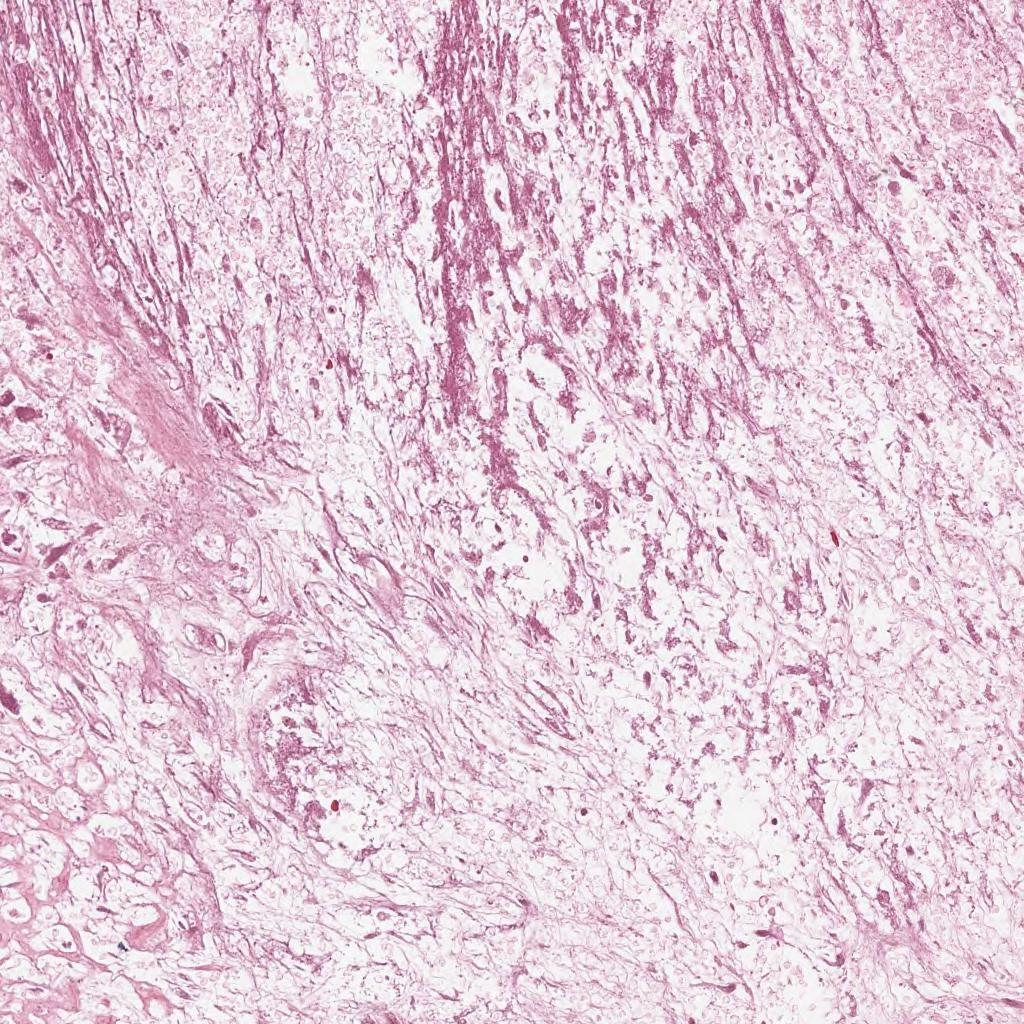}
\caption{Non-viable Tumor}
\end{subfigure}
\begin{subfigure} [b]{0.3\textwidth}
\includegraphics[width=35mm]{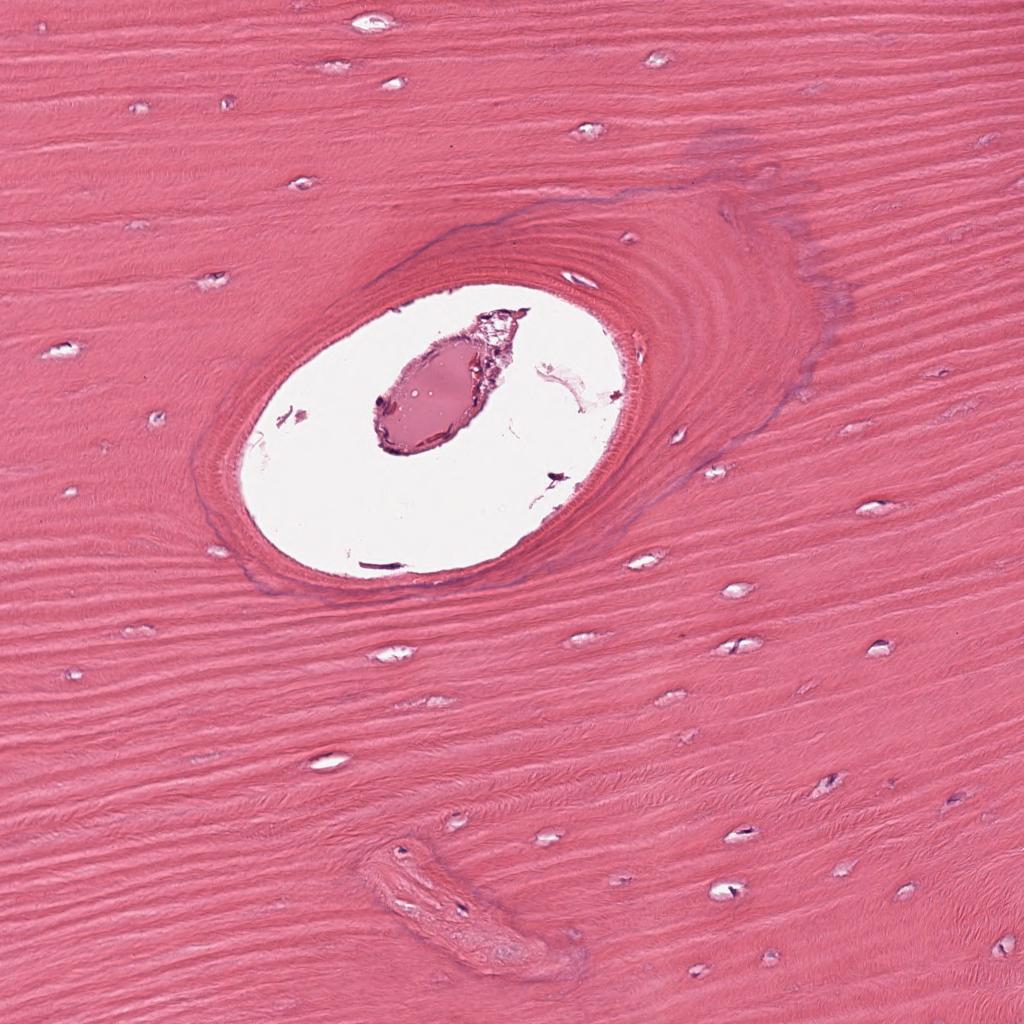}
\caption{Non-Tumor}
\end{subfigure}
\begin{subfigure} [b]{0.3\textwidth}
\includegraphics[width=35mm]{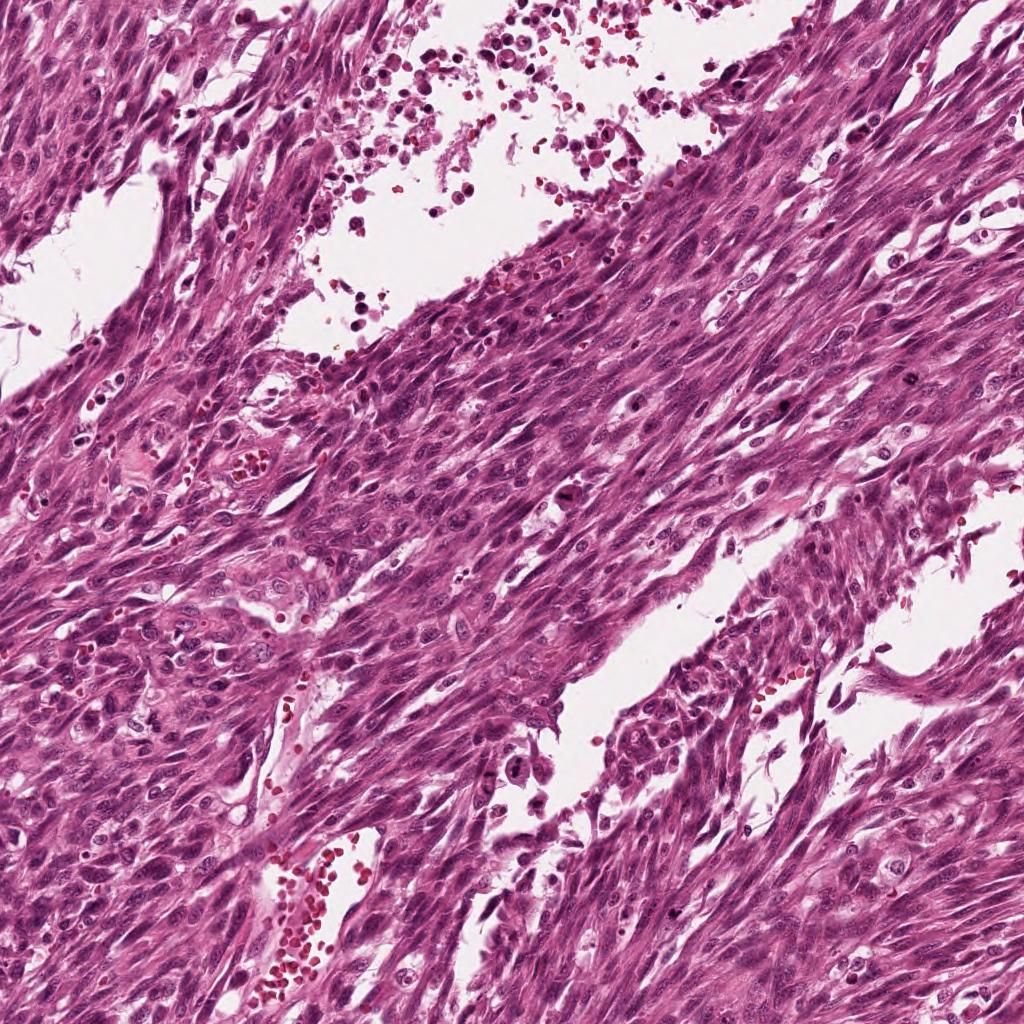}
\caption{Viable Tumor}
\end{subfigure}
\caption{Types of Osteosarcoma Images- A. Non-viable Tumor, B. Non-Tumor, C. Viable Tumor}
\end{figure}

\subsection{Overview of Applied Model}
\subsubsection{EfficientNet}
EfficientNet is the brainchild of the Google Brain Team which studied network scaling and discovered that the performance of a TL model may be improved by adjusting network depth, width, and resolution. They scaled a neural network to build more deep learning models, which have far higher efficacy and accuracy than the CNNs that were previously in use. EfficientNet produced accurate and reliable large-scale visual recognition for the ImageNet. EfficientNet utilizes a composite scaling technique where the network depth is corresponding to the number of layers in a network. The width of a convolutional layer is related to how many filters are present while the resolution is determined by the input image's height and breadth. EfficientNet showed excellent results even with the transfer learning method, proving its usefulness outside of the ImageNet dataset. There are 8 versions of the model is available based on parameters and accuracy. This study applies EfficientNetB7 for tumour detection \cite{Marq2020}.

\subsubsection{Resnet}
ResNet, also known as deep residual network, was introduced in 2016 as a solution to the problems associated with DL training such as lack of layers and being time-consuming. This complex model was developed to provide a shortcut to the process without damaging the performance. Additionally, it is now possible to train networks more effectively as computation calculations are made lighter. The model skips ReLu-activated layers along with batch normalization. It also uses Stochastics Gradient Descent (SGD) to set weights and parameters. ResNet has a reputation for outperforming other models in terms of feature extraction \cite{Sarwinda2021}.

\subsubsection{InceptionResNet}
InceptionResNet is a popular TL model made by combining two of the most efficient models Inception and ResNet. Trained on the ImageNet database, this model uses batch normalization instead of convolutional layer summation. Input units are randomly set to 0 in the dropout layers in the training phase to avoid overfitting. The model also uses a flattering technique to create a one-dimensional data array before using it in the following layers. The same process is applied to output data to create a feature vector. The model builds a fully connected layer by connecting to the final model for classification and a "binary cross-entropy" loss function is used along with a batch size of 32 \cite{Faruk2021}.

\subsubsection{Nasnet}
Due to its enormous computing power and technical expertise, Google created the Neural Architecture Search Network-Large (NASNet-Large), which classified the challenge of discovering the perfect CNN architecture as a Reinforcement Learning (RL) problem. NasNet introduces a machine-assisted neural network architectural design which builds a deep neural network without the underlying model which is different from models focusing on tensor decomposition or quantization. The three main operations in contemporary NAS architectures are training a super-network, analysing sampled networks, and training the found network. The compensation for each search cycle was the accuracy of the searched architecture on the provided dataset. NASNet-Large achieved a cutting-edge result in the ImageNet competition. There is a specific image input size for the model, 331 x 331 which is not changeable \cite{Mavra2022}.

\subsection{Evaluation Criteria}
The output is assessed on 5 criteria: accuracy, loss, precision, recall and AUC. Outputs are divided into 4 categories: True Positive (TP), True Negative (TN), False Positive (FP), and False Negative (FN). The categories are explained in table 1:

\begin{table}
    \centering
    \caption{Explanation of TP, TN, FP, and FN}
    \vspace{10pt}
\begin{tabular}{||c | c | c||} 
 \hline
 I/O & Output Positive & Output Negative \\ [0.5ex] 
 \hline\hline
 Input Positive & True Positive (TP) & False Positive (FP) \\ 
 \hline
 Input Negative & False Negative (FN) & True Negative (TN) \\ [1ex] 
 \hline
\end{tabular}
\end{table}

\subsubsection{Accuracy}
Expressed by percentage, accuracy measures how well the predicted outcome matches the actual outcome. It is calculated by multiplying the total number of possible outcomes by the number of true positive and true negative outcomes:

Accuracy = (TP + FN) / (TP + TF + FP + FN)

\subsubsection{Precision}
Precision is calculated by dividing the true positive by the total of true and false positives. It indicates the degree to which projected outcomes agree with one another.

Precision = TP / (TP + FP)

\subsubsection{Recall}
By dividing the total number of true positives by the sum of true positives and false negatives, Recall is calculated:

Recall = TP / (TP + FN)

The Adam Optimizer was used for the experiment. The name ‘Adam’ was derived from ‘Adaptive Moment Estimation’. Because of its property that combines the efficiency of AdaGrad for sparse gradients with the ability of RMSProp to perform effectively in non-stationary environments, Adam has become one of the most popular optimizers used for deep learning \cite{Soy2020}. 

Other parameters used in this study for the selected models are mentioned in Table 2:

\begin{table}
    \centering
    \caption{Parameters used to build the selected models} 
    \vspace{10pt}
\begin{tabular}{||c | c||} 
 \hline
 Parameter & Value \\ 
 \hline\hline
 Batch size & 32 \\ 
 \hline
 Epochs & 50 \\ 
 \hline
  Runtime & GPU \\ 
 \hline
  Platform & Google Colab \\ 
\hline
  Loss & Categorical Loss Entropy \\ 
  \hline
\end{tabular}
\end{table}

\section{Result Analysis}

\begin{table}
    \centering
    \caption{Performance of the selected models for Osteosarcoma Detection}
    \vspace{10pt}
\begin{tabular}{||c | c| c | c | c | c ||} 
 \hline
 Model & Loss & Accuracy & Precision & Recall & AUC \\  
 \hline\hline
 For Training Set \\
 \hline
 EfficientNetB7 & 40.6399 & 71.69\% & 0.4337 & .4337 & 0.6309 \\ 
 \hline
  InceptionResNetV2 & 0.5040 & 98.30\% & 0.9660 & 0.9660 & 0.9830 \\ 
 \hline
  NasNetLarge & 1.3211 & 98.80\% & 0.9759 & 0.9759 & 0.9839 \\ 
\hline
  ResNet50 & 0.8564 & 89.16\% & 0.7895 & 0.7722 & 0.9324 \\ 
 \hline
  For Validation Set \\
 \hline
 EfficientNetB7 & 83.9986 & 62.77\% & 0.2554 & .2554 & 0.5036 \\ 
 \hline
  InceptionResNetV2 & 7.3212 & 93.29\% & 0.8658 & 0.8658 & 0.9195 \\ 
 \hline
  NasNetLarge & 16.4955 & 90.91\% & 0.8182 & 0.8182 & 0.8886 \\ 
\hline
  ResNet50 & 1.3083 & 89.83\% & 0.7991 & 0.7992 & 0.9250 \\ 
  \hline
\end{tabular}
\end{table}

From the output described in Table 3, several decisions can be made. First of all, in all cases, the value of the loss is way bigger for the validation set than for the training set. NasNetLarge has the biggest difference between the two values (Loss for validation is 12.48 times larger than the training set). ResNet50 has the least loss values along with the least difference between the values (Loss for validation set is greater or equal than (1.52 x Loss for training set)). For both training and validation sets, the precision and recall value is the same for a model except for ResNet50. InceptionResNetV2 has the highest precision and recall values for the validation set (0.8658) whereas NasNetLarge has the highest for the training set (0.9759). EfficientNetB7 has the lowest precision and recall values among the 4 for both sets.

For the training set, NasNetLarge has the highest accuracy (98.80\%), followed by InceptionResNetV2 (98.30\%), ResNet50 (89.16\%) and EfficientNetB7 (71.69\%). For the validation set, InceptionResNetV2 has the highest accuracy (93.29\%), followed by NasNetLarge (90.91\%), ResNet50 (89.83\%), and EfficientNetB7 (62.77\%). It can be concluded from the result that InceptionResNetV2 and NasNetLarge are the best models for Osteosarcoma detection.

\section{Conclusion}
The use of Transfer Learning for medical image analysis decreases the chance of false detection along with a reduction of complexity, required time, human involvement, and cost which can ensure proper medical treatment for people, even the ones with limited income. The paper studies the efficiency of different transfer learning models for Osteosarcoma tumour detection which is one of the most prominent types of cancer but was not studied enough by researchers who are interested in exploring the applications of transfer learning models in this domain. The output provided by the models was analysed on different criteria to understand them better. Among the 4 models, InceptionResNetV2 and NasNetLarge have provided the best performance. ResNet50 can also be considered an efficient model for tumour detection, however, the performance provided by EfficientNetB7 was not satisfactory. This study indicates further research opportunities such as:

\begin{itemize}
    \item A comparison among other transfer learning models to study if any other model can outperform these 4 models.
    \item A further study to increase the accuracy of EfficientNetB7 for Osteosarcoma Detection.
    \item Though InceptionResNetV2 and NasNetLarge achieved the highest accuracies, the study can be further conducted to improve them further.
    \item The models can be used for other types of bone disease detection and diseases in other organs such as bone marrow.
\end{itemize}

\end{document}